# Prior Frequency Guided Diffusion Model for Limited Angle (LA)-CBCT Reconstruction

## Running Title: PFGDM for LA-CBCT Reconstruction

Jiacheng Xie

Hua-Chieh Shao

Yunxiang Li

You Zhang

*The Advanced Imaging and Informatics for Radiation Therapy (AIRT) Laboratory*
*The Medical Artificial Intelligence and Automation (MAIA) Laboratory*
*Department of Radiation Oncology, University of Texas Southwestern Medical Center, Dallas, TX 75390, USA*

Corresponding address:

You Zhang
Department of Radiation Oncology
University of Texas Southwestern Medical Center
2280 Inwood Road
Dallas, TX 75390
Email: You.Zhang@UTSouthwestern.edu
Tel: (214) 645-2699

## Abstract

*Objective.* Cone-beam computed tomography (CBCT) is widely used in image-guided radiotherapy. Reconstructing CBCTs from limited-angle acquisitions (LA-CBCT) is highly desired for improved imaging efficiency, dose reduction, and better mechanical clearance. LA-CBCT reconstruction, however, suffers from severe under-sampling artifacts, making it a highly ill-posed inverse problem. Diffusion models can generate data/images by reversing a data-noising process through learned data distributions; and can be incorporated as a denoiser/regularizer in LA-CBCT reconstruction. In this study, we developed a diffusion model-based framework, prior frequency-guided diffusion model (PFGDM), for robust and structure-preserving LA-CBCT reconstruction. *Approach.* PFGDM uses a conditioned diffusion model as a regularizer for LA-CBCT reconstruction, and the condition is based on high-frequency information extracted from patient-specific prior CT scans which provides a strong anatomical prior for LA-CBCT reconstruction. Specifically, we developed two variants of PFGDM (PFGDM-A and PFGDM-B) with different conditioning schemes. PFGDM-A applies the high-frequency CT information condition until a pre-optimized iteration step, and drops it afterwards to enable both similar and differing CT/CBCT anatomies to be reconstructed. PFGDM-B, on the other hand, continuously applies the prior CT information condition in every reconstruction step, while with a decaying mechanism, to gradually phase out the reconstruction guidance from the prior CT scans. The two variants of PFGDM were tested and compared with current available LA-CBCT reconstruction solutions, via metrics including peak signal-to-noise ratio (PSNR) and structure similarity index measure (SSIM). *Main Results.* PFGDM outperformed all traditional and diffusion model-based methods. The mean(s.d.) PSNR/SSIM were 27.97(3.10)/0.949(0.027), 26.63(2.79)/0.937(0.029), and 23.81(2.25)/0.896(0.036) for PFGDM-A, and 28.20(1.28)/0.954(0.011), 26.68(1.04)/0.941(0.014), and 23.72(1.19)/0.894(0.034) for PFGDM-B, based on 120°, 90°, and 30° scan angles respectively. In contrast, the PSNR/SSIM was 19.61(2.47)/0.807(0.048) for 30° for DiffusionMBIR, a diffusion-based method without prior CT conditioning. *Significance.* PFGDM reconstructs high-quality LA-CBCTs under very-limited gantry angles, allowing faster and more flexible CBCT scans with dose reductions.

**Keywords: Cone-beam CT, diffusion model, limited angle, image reconstruction**

## 1. Introduction

Cone-beam computed tomography (CBCT) has been widely applied to fields like radiotherapy for efficient and high-resolution volumetric imaging (Walter *et al.*, 2020). Limited-angle CBCT (LA-CBCT), which allows faster imaging speed, lower X-ray dose, and better mechanical clearance, has been investigated extensively (Zhang *et al.*, 2013a; Zhang *et al.*, 2013b; Je *et al.*, 2014; Ren *et al.*, 2014; Zhang *et al.*, 2015b). The analytical filtered back-projection-based Feldkamp-Davis-Kress (FDK) (Feldkamp *et al.*, 1984) algorithm is widely used for CBCT reconstruction. However, given limited scan angles for LA-CBCT, FDK produces low-quality reconstructions with under-sampling artifacts and anatomical distortions. A well-established alternative for tomographic reconstruction is model-based iterative reconstruction (MBIR) (Katsura *et al.*, 2012; Liu, 2014), which formulates the reconstruction as a 3D inverse problem and solves it by optimizing a physics-based data consistency model with additional regularization terms. For regularization, the total variation (TV) penalty is widely used due to its edge-preserving and smoothing capabilities (Sidky and Pan, 2008). MBIR or similar techniques have achieved substantial reconstruction improvement for sparsely sampled or low-dose projections (Katsura *et al.*, 2012; Pickhardt *et al.*, 2012; Chang *et al.*, 2013; Rui *et al.*, 2015). However, in the context of LA-CBCT reconstruction, these methods usually produce sub-optimal results given the lack of information from a full block of scan angles, which necessitates the use of stronger imaging priors to guide the reconstruction. To introduce patient-specific priors, registration-driven approaches, which use limited-angle projections to solve LA-CBCTs via deforming previously acquired CT/CBCT images (Zhang *et al.*, 2013b; Ren *et al.*, 2014; Zhang, 2021), have been introduced. However, compared with direct reconstruction-based methods, the registration-driven methods may fall short in solving non-deformation-induced anatomy changes (Zhang *et al.*, 2017a), and the imaging intensity variations between different imaging systems/protocols may yield such methods less stable (Zhang *et al.*, 2015a).

With the recent success of deep learning (DL) techniques, several works have been proposed for limited-angle/view image reconstruction. For instance, sinogram-based methods used generative adversarial networks (GANs) to predict full-angle/view projection data from limited-angle/view acquisitions, and then employed traditional methods like FDK to reconstruct artifact-free CBCTs from the predicted fully sampled data (Li *et al.*, 2019). Image-domain methods, on the other hand, employed convolutional neural networks (CNNs) to enhance image quality by directly mapping the artifacts-ridden reconstructions to corresponding high-quality images (Chen *et al.*, 2017; Gu and Ye, 2017; Gupta *et al.*, 2018). Meanwhile, dual-domain methods such as DuDoNet (Lin *et al.*, 2019) trained CNNs in both sinogram space and image space to improve the image reconstruction quality and outperformed the former two strategies. However, such methods are often trained under a fixed under-sampling geometry/scenario, and may not generalize well to scenarios with flexible under-sampling patterns. Methods that combine neural networks with iterative physics-informed reconstruction models can often achieve better results and are more robust to scan scenario variations (Chen *et al.*, 2022; Hu *et al.*, 2022). These hybrid methods effectively leverage the power of neural networks in feeding prior information and serving as denoisers, and the proven records of physics-driven iterative reconstructions in solving model-based inverse problems. Different neural networks, including GAN, have been used as imaging priors to regularize model-based reconstructions, including LA-CBCT reconstruction (Yang *et al.*, 2020; Barutcu *et al.*, 2021; Zhou *et al.*, 2021; Gao *et al.*, 2023). However, such methods may still fail, especially for very small scan angles, as the priors provided by these neural networks are insufficient or unstable.

Recently, diffusion models were introduced into the inverse problems of image generation/reconstruction, and shown to outperform popular generative models such as GAN (Dhariwal and Nichol, 2021). Diffusion models operate through forward and backward diffusion steps (Ho *et al.*, 2020). In the forward process, a diffusion model gradually adds noise to an image, transforming the data into a Gaussian noise distribution. Starting from the noise, the model then iteratively reduces the noise, through a reverse diffusion process which is trained to reconstruct the original image or generate new samples from the learned data distribution. In the field of medical imaging, score-based diffusion models (Song and Ermon, 2019; Song and Ermon, 2020; Song *et al.*, 2020b; Song *et al.*, 2021) are a class of methods that achieve state-of-the-art (SOTA) performance in image generation. It provides a powerful way to learn population-based imaging statistics and distributions and can serve as strong imaging priors in medical imaging studies (Song *et al.*, 2021; Chung and Ye, 2022). In terms of limited-angle CT (LA-CT) reconstruction, DiffusionMBIR (Chung *et al.*, 2023) and DOLCE (Liu *et al.*, 2022) are among the latest diffusion model-based methods that showed superior performance over the previous methods. By embedding the diffusion model as a denoiser in an alternating direction method of multipliers (ADMM)-based iterative reconstruction framework, DiffusionMBIR uses the prior distribution learned in the diffusion model to enhance the LA-CT image quality. DOLCE (Liu *et al.*, 2022), on the other hand, similarly used the diffusion model to regularize an iterative model-based reconstruction, while conditioning the diffusion model with additional LA-CTs reconstructed by regularized least squares (RLS) or filtered back-projection (FBP) techniques. However, both the DOLCE technique and the DiffusionMBIR technique were developed for CT reconstruction using parallel beam geometries. A clinically realistic cone-beam imaging geometry can be more challenging for both methods. In addition, although both methods showed improved results in limited-angle reconstruction, it is still challenging to reconstruct high-quality and accurate images using relatively small gantry angles (i.e., $\theta \leq 90°$) due to the lack of patient-specific information.

To further explore the potential of diffusion model-driven reconstruction, we proposed a prior frequency-guided diffusion model (PFGDM) for LA-CBCT reconstruction. PFGDM is built upon the findings of our recent work on zero-shot image translation (Li *et al.*, 2023), which observed that a diffusion model conditioned on image-domain high-frequency information can better retain the anatomical details within the images. For source and target image domains of the same or similar underlying anatomy (for instance, CBCT and CT image domains of the same patient), the high-frequency information from one source image domain can be applied as patient-specific priors to condition the diffusion model for structural preservation during source-to-target domain image translation. In the context of radiotherapy, it is almost certain that patient-specific CTs of relatively high quality will be acquired for treatment simulation/planning before the following CBCT acquisitions for treatment guidance. Such CT information can serve a strong anatomical prior to condition the reconstruction of daily LA-CBCTs. Inspired by these observations, we developed PFGDM for LA-CBCT reconstruction, which is conditioned on high-frequency information extracted from patient-specific prior CT scans. However, although the anatomical structures are expected to be similar between CT/CBCT, they may not be exactly the same due to daily anatomical variations. To allow the reconstruction of CBCT anatomical structures that are different from those of CT, we proposed two ways of introducing the prior CT condition to the reconstruction: 1) PFGDM-A, which uses a condition-dropping mechanism to drop the prior CT condition after a preset step number, and continue reconstructing the CT/CBCT anatomical differences in the remaining reconstruction steps. 2) PFGDM-B, which continuously applies the conditional input in every reconstruction step. However, in each step, an increasing high-pass threshold was applied to the prior CT to gradually remove the edge information of the anatomical structures and phase out the guidance from the prior CT.

In this study, we evaluated the two variants of PFGDM and compared them with the other methods for LA-CBCT reconstructions across a range of limited angles ($\theta \in 30°, 90°, 120°$) and three anatomical sites (lung, pelvis, and head and neck). We also evaluated the PFGDM's potential on CBCT reconstruction using extremely limited scan angles (2° to 10°). Trained as a single model, PFGDM achieves SOTA performance across all scenarios, making it a highly promising tool for clinical applications.

## 2. Materials and Methods

### 2.1 Background of LA-CBCT reconstruction and related methods

**Inverse problems.** Reconstructing LA-CBCT can be formulated as a linear inverse problem:

$$y = Ax + n, \tag{1}$$

where $x \in \mathbb{R}^n$ is the image to be reconstructed, $y \in \mathbb{R}^m$ is the measurement (i.e., sinogram), $A \in \mathbb{R}^{m \times n}$ is the measurement system matrix, and $n$ is the measurement noise in the system. For CBCT reconstruction, a standard optimization-based approach usually iteratively estimates $\boldsymbol{x}$ from the measurement $\boldsymbol{y}$ with additional TV regularization:

$$\boldsymbol{x}^* = \arg\min_{\boldsymbol{x}} \left( \frac{1}{2} ||\boldsymbol{y} - \boldsymbol{A}\boldsymbol{x}||_2^2 + ||\boldsymbol{D}\boldsymbol{x}||_1 \right), \tag{2}$$

where $\boldsymbol{D} := [\boldsymbol{D_x}, \boldsymbol{D_y}, \boldsymbol{D_z}]^T$ represents the finite difference operator in each axis. Minimization of Eq. 2 can be performed with optimization algorithms such as ADMM. Reconstructing $\boldsymbol{x}$ from $\boldsymbol{y}$ is highly ill-posed for LA-CBCT, thus additional assumptions/priors on $\boldsymbol{x}$ are often needed.

**Score-based diffusion models.** Score-based diffusion models (Song *et al.*, 2020b) are a type of generative models that simulate the generation of data/image by reversing a data-noising process, which can be incorporated as a denoiser in inverse problems. Consider a stochastic process with the time variable $t \in [0,1]$, the noise-corrupted data/image $\boldsymbol{x}$ can be represented as $\boldsymbol{x}(t) = \boldsymbol{x}_t$. Suppose the data is sampled from a distribution $p(\boldsymbol{x})$ for $\boldsymbol{x}(0)$, as time progresses ( $t = 1$ ), the data distribution $\boldsymbol{x}(1)$ approaches a Gaussian distribution. The evolution of the data distribution over time is described by a Stochastic Differential Equation (SDE):

$$d\boldsymbol{x} = \boldsymbol{f}(\boldsymbol{x}, t)dt + g(t)d\boldsymbol{w}, \tag{3}$$

where $\boldsymbol{f}(\boldsymbol{x}, t): \mathbb{R}^{n \times 1} \mapsto \mathbb{R}^n$ is a drift function, $g(t): \mathbb{R} \mapsto \mathbb{R}$ is a scalar diffusion function, and $\boldsymbol{w}$ denotes the $n$- dimensional standard Brownian motion (Särkkä and Solin, 2019). When $\boldsymbol{f}(\boldsymbol{x}, t) = 0$ and $g(t) = \frac{\sqrt{d[\sigma^2(t)]}}{dt}$, the SDE simplifies to the Brownian motion:

$$d\boldsymbol{x} = \frac{\sqrt{d[\sigma^2(t)]}}{dt} d\boldsymbol{w}, \tag{4}$$

which is a specific case of SDE: Variance Exploding SDE (VE-SDE) (Song *et al.*, 2020b). The data evolution in VE-SDE is characterized by a constant mean with increasing Gaussian noise, eventually leading to pure Gaussian noise. By applying Anderson's theorem (Anderson, 1982), the reverse of this SDE, which forms the basis of the generative process in score-based diffusion models, is defined as (Song *et al.*, 2020b):

$$d\overline{\boldsymbol{x}} = -\frac{d[\sigma^2(t)]}{dt} \nabla_{\boldsymbol{x}_t} \log p\left(\boldsymbol{x}_t\right) dt + \frac{\sqrt{d[\sigma^2(t)]}}{dt} d\overline{\boldsymbol{w}}, \tag{5}$$

where $\nabla_{\boldsymbol{x}_t} \log p\left(\boldsymbol{x}_t\right)$ is the score function, representing the gradient of the log probability with respect to the data. To solve the reverse-time SDE that defines the generative process of the diffusion model, we train the neural network through Denoising Score Matching (DSM) (Vincent, 2011; Song *et al.*, 2020b):

$$\min_{\theta} E_{t, \boldsymbol{x}(t)} \left[ \lambda(t) || \boldsymbol{s}_{\theta}(\boldsymbol{x}(t), t) - \nabla_{\boldsymbol{x}_t} \log p\left(\boldsymbol{x}(t) | \boldsymbol{x}(0)\right) ||_2^2 \right], \tag{6}$$

where $\boldsymbol{s}_{\theta}(\boldsymbol{x}(t), t): \mathbb{R}^{n \times 1} \mapsto \mathbb{R}^n$ is a time-dependent neural network and $\lambda(t)$ is the weighting function. Ultimately, given this robust training schema (Chung *et al.*, 2023), $\boldsymbol{s}_{\theta^*}(\boldsymbol{x}(t), t) \simeq \nabla_{\boldsymbol{x}_t} \log p\left(\boldsymbol{x}_t\right)$ can be established. Hence, the reverse diffusion process is approximated by:

$$d\overline{\boldsymbol{x}} \simeq -\frac{d[\sigma^2(t)]}{dt} \boldsymbol{s}_{\theta^*}(\boldsymbol{x}(t), t) dt + \sqrt{\frac{d[\sigma^2(t)]}{dt}} d\overline{\boldsymbol{w}}. \tag{7}$$

To solve Eq. 7, one can employ methods such as the predictor-corrector (PC) sampler (Song *et al.*, 2020b), which involves discretizing the time interval from 0 to 1 into $N$ equal steps. It allows the numerical integration of the reverse diffusion process to generate samples from the target distribution.

**Diffusion model in inverse problems.** Training diffusion models involves solving the reverse SDE with the approximated score function defined in Eq. 7, which can be viewed as sampling from the prior distribution $p(\boldsymbol{x})$. To solve inverse problems, we need to sample from the posterior distribution $p(\boldsymbol{x}|\boldsymbol{y})$. Based on the Bayes' rule:

$$p(\boldsymbol{x}|\boldsymbol{y}) = \frac{p(\boldsymbol{x})p(\boldsymbol{y}|\boldsymbol{x})}{p(\boldsymbol{y})}. \tag{8}$$

DiffusionMBIR (Chung *et al.*, 2023) is the one of the representative work that introduced diffusion models into solving inverse problems of reconstructing limited-angle CT images. In DiffusionMBIR's framework, given the definition in Eq. 1, the inverse problem was solved using the diffusion prior in Eq. 7 with the Bayes' rule, leading to:

$$\begin{aligned}
\nabla_{\boldsymbol{x}_t} \log p\left(\boldsymbol{x}_t|\boldsymbol{y}\right) &= \nabla_{\boldsymbol{x}_t} \log p\left(\boldsymbol{x}_t\right) + \nabla_{\boldsymbol{x}_t} \log p\left(\boldsymbol{y}|\boldsymbol{x}_t\right) \\
&\simeq \boldsymbol{s}_{\theta^*}(\boldsymbol{x}(t), t) + \nabla_{\boldsymbol{x}_t} \log p\left(\boldsymbol{y}|\boldsymbol{x}_t\right).
\end{aligned} \tag{9}$$

To incorporate Eq. 9, the existing literature (Song *et al.*, 2020b; Chung *et al.*, 2022) defined the update step as:

$$\boldsymbol{x}'_{i-1} \leftarrow \text{Solve}(\boldsymbol{x}_{i-1}, \boldsymbol{s}_{\theta^*}), \tag{10}$$
$$\boldsymbol{x}_{i-1} \leftarrow \mathcal{P}_{\{\boldsymbol{y} = \boldsymbol{Ax} + \boldsymbol{n}\}}(\boldsymbol{x}'_{i-1}), \tag{11}$$

where $Solve$ denotes the reverse-SDE solver of Eq. 7, and $\mathcal{P}_C$ denotes the projection operator to the set $C = \{\boldsymbol{x}|\boldsymbol{y} = \boldsymbol{Ax} + \boldsymbol{n}\}$.

**High-frequency information extraction.** To extract the high-frequency information from prior CT images for PFGDM reconstruction, we used the Sobel operator (Kanopoulos *et al.*, 1988), a discrete differential operator. Specifically, we convolved the CT image $CT_{prior}$ with horizontal and vertical Sobel filter kernels, denoted as $K_x$ and $K_y$, respectively, resulting in two filtered images, $G_x$ and $G_y$. The gradient magnitude is obtained by combining the resulting gradient approximations:

$$G_x = K_x * CT_{prior}, \quad G_y = K_y * CT_{prior}, \tag{12}$$
$$i_{\text{mag}} = \sqrt{G_x^2 + G_y^2}. \tag{13}$$

The high-frequency information $H$ can then be extracted through applying a threshold $\eta$:

$$H_\eta = \begin{cases} i_{\text{mag}}(x, y) & i_{\text{mag}}(x, y) \geq \eta \\ 0 & i_{\text{mag}}(x, y) < \eta \end{cases}. \tag{14}$$

*2.2 The proposed method: PFGDM*

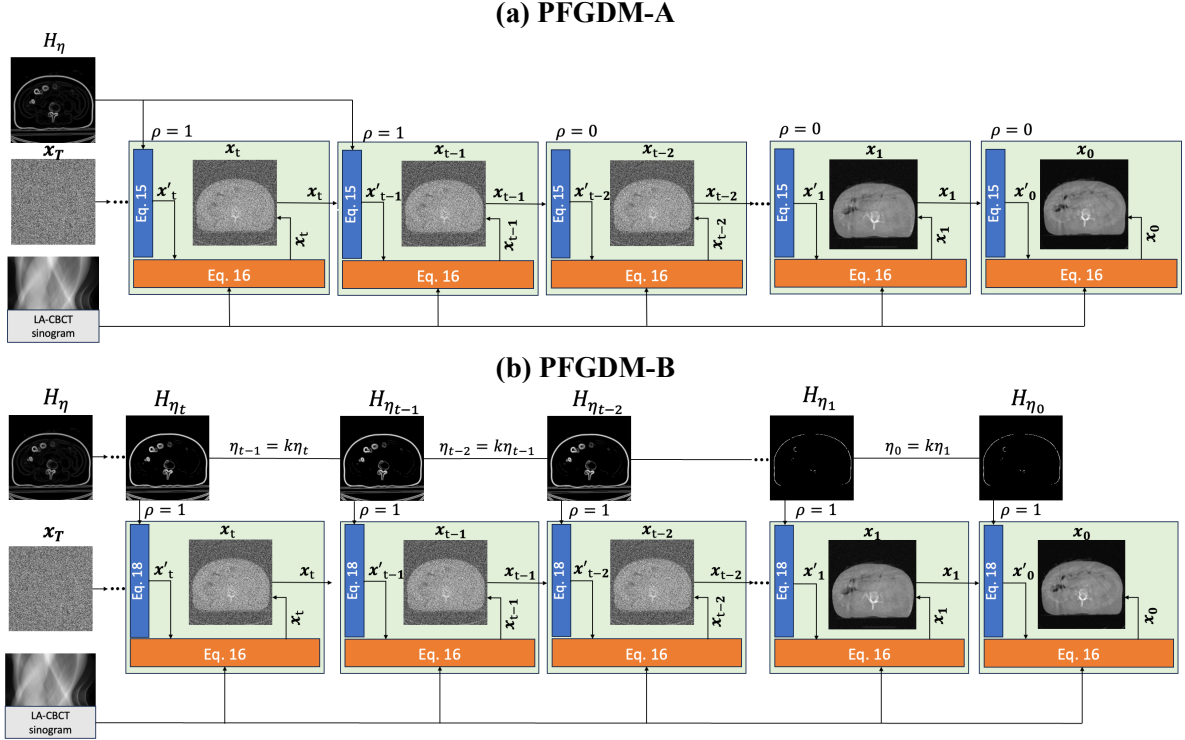

**Figure 1.** Overview of the two PFGDM variants: PFGDM-A and PFGDM-B. Starting from the Gaussian noise $\boldsymbol{x}_T$, in each SDE step of the diffusion model (blue box), we sample a LA-CBCT image $\boldsymbol{x}_t$ from the posterior by applying the reverse diffusion that is conditioned on the high-frequency prior CT information. Concurrent with the diffusion model, the LA-CBCT image is also iteratively updated via ADMM based on the given LA-CBCT sinogram (orange box), until convergence. (a) For PFGDM-A, the conditional input $H_\eta$ remains unchanged during reconstruction and is dropped after step $t-1$. (b) For PFGDM-B, the conditional input $H_{\eta_t}$ is continuously introduced along the process while dynamically updated to gradually phase out the edge information of the prior CT.

An overview of the PFGDM method and its two variants is shown in Fig. 1. In general, PFGDM reconstructs CBCTs from LA-CBCT projections with the guidance of a conditional diffusion model, where the condition $H$ is obtained by applying the Sobel operator on the prior CT (Eq. 14). Solving the inverse problem with the diffusion model involves solving the reverse SDE with the approximated score function defined in Eq. 7, which can be viewed as sampling from the conditional posterior distribution $p(\boldsymbol{x}|\boldsymbol{y}, H)$. To avoid the excessive computational hardware/power required by 3D diffusion models, we use the 2D diffusion model for 3D LA-CBCT reconstruction. Inspired by DiffusionMBIR (Chung *et al.*, 2023), we applied the diffusion model slice by slice while adding TV regularization along the z-axis (slice direction) in parallel to regularize the inter-dependency between slices. Specifically, Eq.10 is enforced by applying a 2D diffusion model-based regularizer slice-by-slice, while Eq. 11 is realized through ADMM-based iterative reconstruction with TV regularization.

Thus, the solution of PFGDM-A can be formulated as:

$$\boldsymbol{x}'_{i-1} \leftarrow \text{Solve}\big(\boldsymbol{x}_i, \boldsymbol{s}_{\theta^*}, \rho H_\eta\big), \tag{15}$$

$$\boldsymbol{x}_{i-1} \leftarrow \arg\min_{\boldsymbol{x}'_{i-1}} \Big(\frac{1}{2}||\boldsymbol{y} - \boldsymbol{A}\boldsymbol{x}'_{i-1}||^2_2 + ||\boldsymbol{D}_z\boldsymbol{x}'_{i-1}||_1\Big), \tag{16}$$

where $||D_z x'_{i-1}||_1$ represents the L1 norm of the finite differences of $x'_{i-1}$ along the z-axis only, and $i_{mag}$ denotes the Sobel-filtered prior CT information. As the reconstruction is coupled with the reverse diffusion process, the iteration was denoted in a descending order, i.e., from T to 0. Specifically, Eq. 15 equals to denoising each slice in parallel by the trained score-based diffusion model under the conditional input $H_\eta$. Here *Solve* refers to the predictor-corrector solver that alternates between the numerical SDE solver and Monte Carlo Markov chain (MCMC) steps, which shows outstanding performance according to Song's work (Song *et al.*, 2020b). Eq. 16 applies the z-directional TV prior and enforces data fidelity. Notably, $\rho$ is a hyper-parameter that determines whether the conditional input $H_\eta$ exists or not, which controls the condition-dropping mechanism. The conditional input $H_\eta$ of PFGDM-A remains unchanged during the reconstruction process, while $\rho$ is changed from 1 to 0 after a pre-set step number to stop introducing the condition.

For PFGDM-B, the corresponding formulation for the conditional diffusion is:

$$H_{\eta_{i-1}} = \begin{cases} i_{mag} & i_{mag} \geq \eta_{i-1} \\ 0 & i_{mag} < \eta_{i-1} \end{cases}, \quad \eta_{i-1} = \frac{i}{i-1}\eta_i \tag{17}$$

$$x'_{i-1} \leftarrow \text{Solve}(x_i, s_{\theta^*}, \rho H_{\eta_{i-1}}), \tag{18}$$

where $\eta_{i-1} = k\eta_i$ defines the adaptive high-pass threshold. In this reconstruction schema, the increasing high-frequency threshold gradually phases out the edge information along the reconstruction process. Compared with PFGDM-A, PFGDM-B does not need to specifically optimize the step number where the condition is dropped ($\rho = 1$).

## *2.3 Experimental setup*

### 2.3.1 Datasets and pre-processing

| Prior CT | Sobel-filtered CT | CBCT |
|---|---|---|

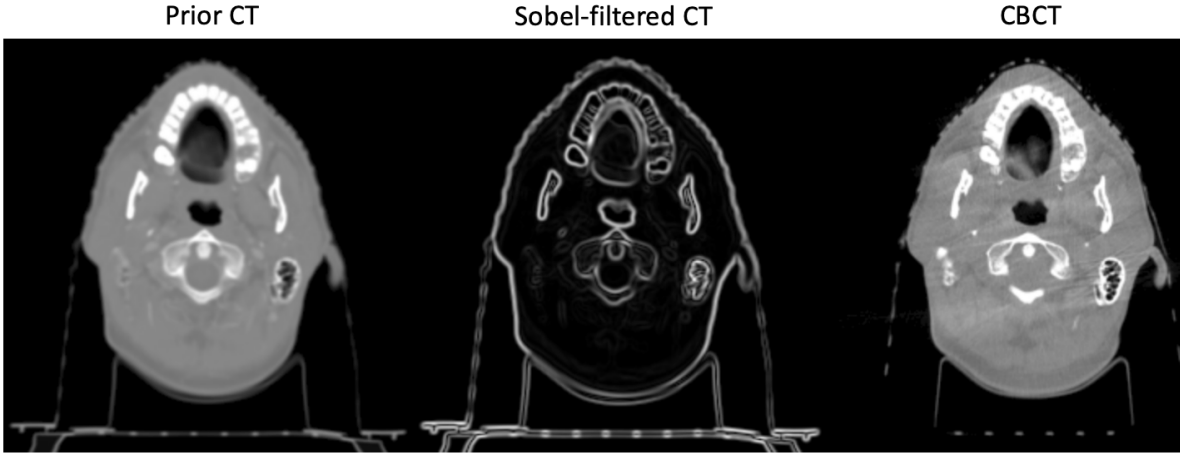

**Figure 2.** Samples from the head and neck dataset: the prior CT scan (left), the extracted high-frequency information of the prior CT scan (middle), and the CBCT scan (right) of the same patient.

To train the conditional score-based diffusion model, we used an in-house dataset consisting of ~3000 CBCT slices from 30 patients of three anatomical sites: head and neck, lung, and pelvis. All CBCT slices were resized to $256 \times 256$. Each CBCT volume was then intercepted by a Hounsfield unit range of [-1000, 1000] and linearly normalized to [0, 1]. The high-frequency information of each CBCT slice is extracted using the Sobel filter function from the Scipy package (Virtanen *et al.*, 2020).

For testing, we used CT and CBCT scans from 12 different patients of the three anatomical sites (4 patients for each site). The CT scans were rigidly registered to the corresponding CBCT images using the open-source software package Elastix (Klein *et al.*, 2010). All CT and CBCT images in the testing set were similarly resized, intercepted, and linearly normalized as the training data. The high-frequency information of the prior CTs was extracted using the Sobel filter as well. Fig. 2 shows a set of sample images from the head and neck dataset.

Cone-beam X-ray projections were simulated on-the-fly using the Operator Discretization Library (ODL) package (Kohr and Adler, 2017) during the test-time LA-CBCT reconstruction. We simulated limited-angle acquisitions from a single direction (single-view) or orthogonal directions (orthogonal-view), for the latter has shown better sampling efficiency under the same total scan angle as in previous works (Zhang *et al.*, 2013b). For single-view projections, the projection angles have ranges of $[0°, 30°]$, $[0°, 90°]$, and $[0°, 120°]$ for $30°$, $90°$, and $120°$ scan angle scenarios, respectively. For orthogonal-view projections, the projection angles have ranges of $([0°, 15°], [90°, 105°])$ for $30°$, $([0°, 45°], [90°, 135°])$ for $90°$, and $([0°, 60°], [90°, 150°])$ for $120°$ scan angle scenarios. The projections were simulated of 256×256 pixels, with each pixel measuring 1.552×1.552 mm$^2$ in dimension. The distance from the source to the detector and to the isocenter was set to 1500 mm and 1000 mm, respectively. The angular sampling density was 1 projection per degree.

*2.3.2 Training and evaluation schemes*

In the training process, the Sobel-filtered CBCT slices were fed as optional conditional input to guide the diffusion model (no CT information needed). Specifically, for PFGDM-A, during the diffusion model training, we used a random number generator to guide random introduction of this extracted high-frequency information with a probability of 0.5 during each reverse diffusion step. Thus the model is trained to fit scenarios with and without the conditional input, allowing us to drop the condition when appropriate to reconstruct CT/CBCT differences. For PFGDM-B, the extracted high-frequency information was consistently applied during each reverse diffusion step, while filtered by a random high-pass threshold that ranges from its minimum to maximum values. The two models were implemented using the Pytorch 2.1.2 library with CUDA 12.1 and trained on an NVIDIA RTX 3090 graphic processing unit (GPU) card separately for one week.

We reconstructed LA-CBCTs using five different methods (FDK, ADMM-TV, DOLCE, DiffusionMBIR, and PFGDM), based on the limited-angle acquisition scenarios simulated in Sec. 2.3.1. We adapted DOLCE (Liu *et al.*, 2022) and DiffusionMBIR (Chung *et al.*, 2023) in this study by modifying the projection geometry from parallel beam to cone beam. Compared with DiffusionMBIR and PFGDM, different DOLCE models need to be retrained for different limited-angle acquisition geometries. To compare the image reconstruction performance of each method, we measured the accuracy of the reconstructed images using PSNR (peak signal-to-noise ratio) and structural similarity index measure (SSIM) (Wang *et al.*, 2004) by comparing with the 'ground-truth' CBCT images. To further evaluate the potential of PFGDM, we compared the reconstruction results of PFGDM-A and PFGDM-B under extremely limited scan angles (2°, 4°, 8°, and 10°). All experiments were run on an NVIDIA Tesla V100 GPU card.

# 3. Results

## *3.1 Hyper-parameter analysis for PFGDM-A*

As aforementioned, in Eq. 15 the hyper-parameter ρ controls the presence of the conditional input during reconstruction for PFGDM-A. With ρ, we can choose to stop introducing the high-frequency information after a preset number of steps during reconstruction, to allow differing CT/CBCT anatomies to be better reconstructed. Figure 3 shows reconstructed LA-CBCTs (second row), with the condition extracted from the prior CT dropped at various iteration steps. It can be seen that dropping the condition at an intermediate step (1400) allows the LA-CBCT to inherit similar anatomical features from the prior CT (1400 vs. 500), while better correcting the anatomical mismatches (1400 vs. 2500).

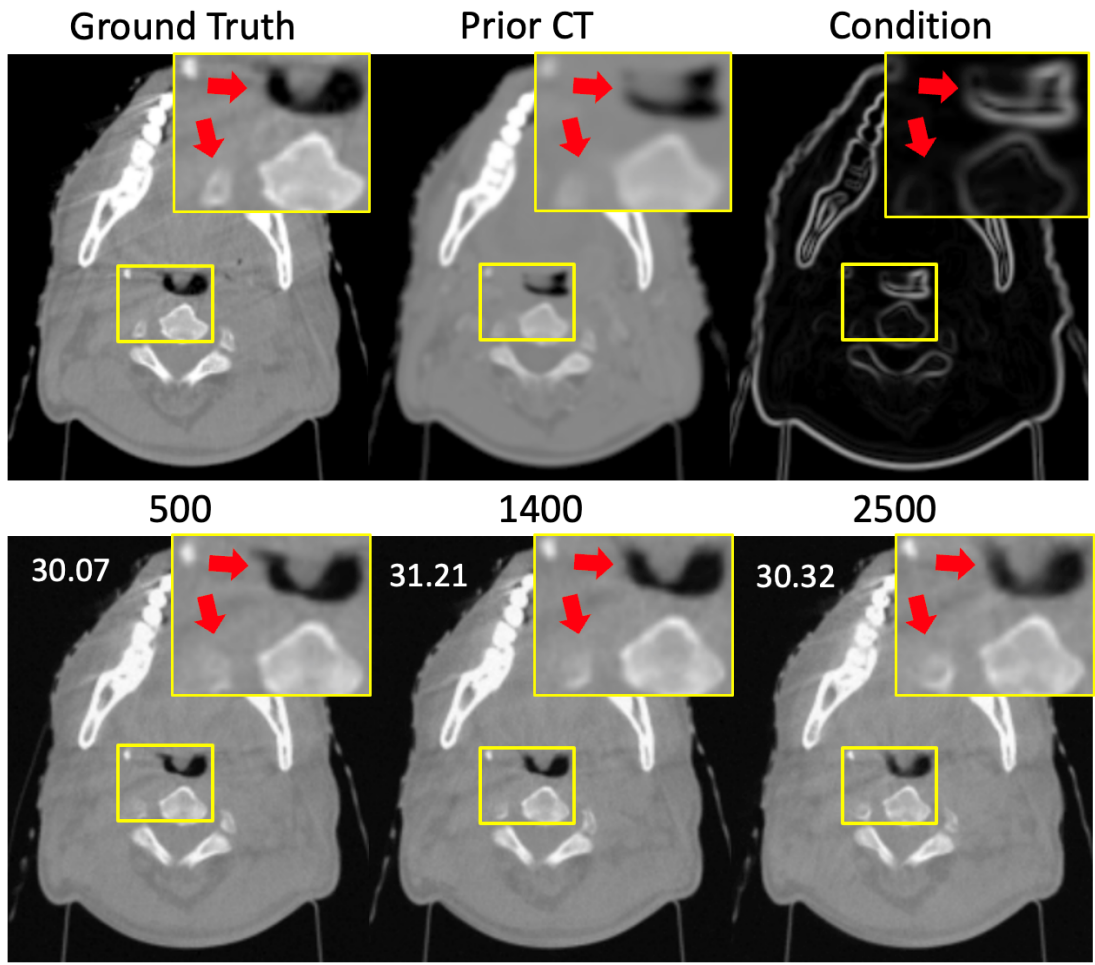

**Figure 3.** Comparison between reconstructed LA-CBCTs, with the high-frequency prior CT condition dropped at different iteration steps (500, 1400, and 2500) for PFGDM-A. The reconstructions are based on an orthogonal-view 90º scan angle (30º for each view), using a total of 3000 iteration steps. PSNR scores measured against the 'ground-truth' fully-sampled CBCT are indicated at the top-left corner.

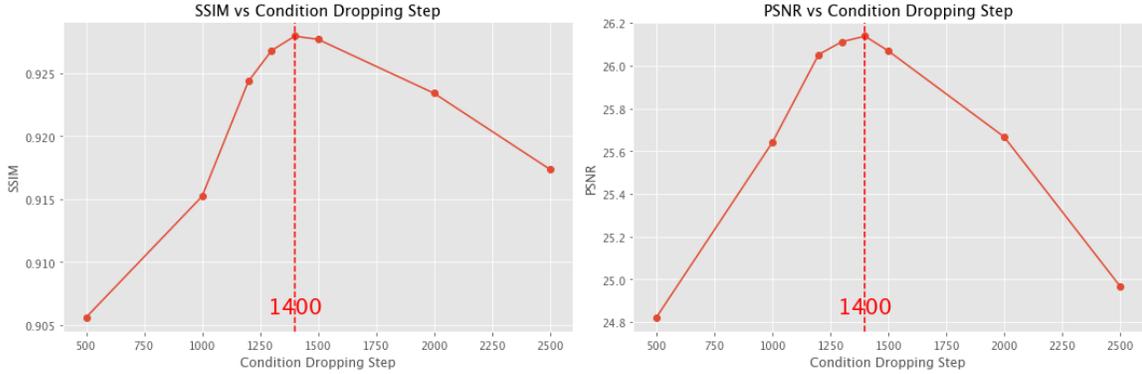

**Figure 4.** Variations of the LA-CBCT reconstruction quality/accuracy given different condition-dropping steps for PFGDM-A on the test set, with a total reconstruction step of 3000. The limited-angle projections were simulated with orthogonal-view acquisitions for a total scan angle of $30°([0°, 15°], [90°, 105°])$, $90°([0°, 45°], [90°, 135°])$, or $120°([0°, 60°], [90°, 150°])$.

To quantitatively determine the optimal step of dropping the high-frequency information for PFGDM-A, we performed a parameterized study using condition-dropping steps ranging from 500 to 2500. We ran the experiments on the test set with the orthogonal-view acquisitions of limited-angle projections for a total scan angle of $30°$, $90°$, and $120°$. The total reconstruction steps for all experiments are 3000. An experiment with the condition dropping step of 500 means that, from step 0 to step 500 the high-frequency condition is introduced to guide the diffusion model, while from step 501 to step 3000 no condition is used. As shown in Fig. 4, the optimal condition-dropping step is around 1400, which gives the best SSIM and PSNR results. All reconstructions performed in the following study for PFGDM-A are based on this step number.

### 3.2 Qualitative evaluation

(a) 120° Total Angle

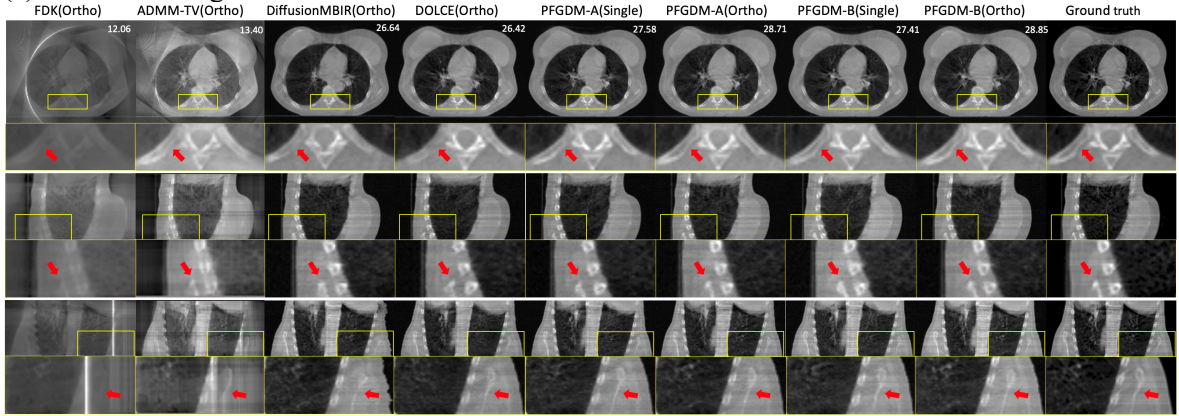

(b) 90° Total Angle

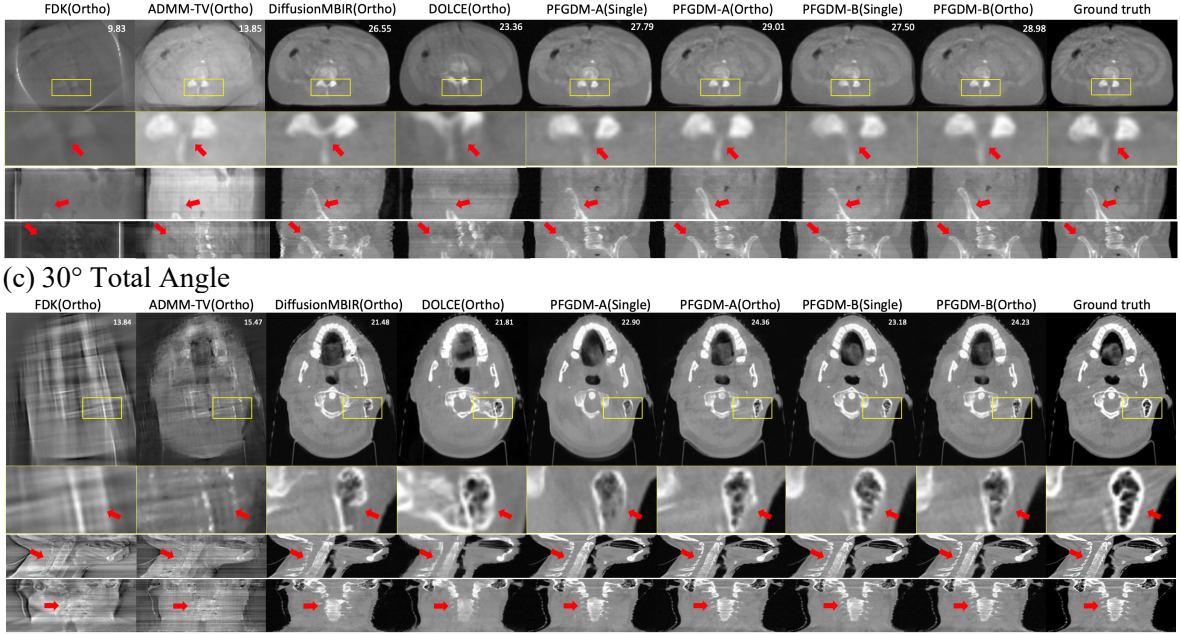

**Figure 5.** Visual comparison between the reconstructed LA-CBCTs of different methods and the 'ground-truth'. In subfigure (a), first, third, and fifth rows represent axial, coronal, and sagittal views, respectively, and the second, fourth, and sixth rows show a zoomed-in region of the corresponding views. In subfigures (b) and (c), first, third, and fourth rows represent axial, coronal, and sagittal views, respectively, and the second row shows a zoomed-in region of the axial view. The limited-angle projections were simulated with either orthogonal-view (ortho) acquisitions for a total scan angle of $30°([0°, 15°], [90°, 105°])$, $90°([0°, 45°], [90°, 135°])$, or $120°([0°, 60°], [90°, 150°])$; or single-view (single) acquisitions for a total scan angle of $30°([0°, 30°])$, $90°([0°, 90°])$, or $120°([0°, 120°])$. Three scenarios of different total scan angles and anatomies are shown as panel a-c respectively. PSNR scores of each reconstruction on axial slices against the 'ground-truth' are indicated at the top-right corner of each figure. See Table I for quantitative results.

We visually compared the reconstructed LA-CBCT images between FDK, ADMM-TV, DOLCE, DiffusionMBIR, PFGDM-A, and PFGDM-B in Fig. 5. Different panels show the results of different scan angles and anatomical sites. The LA-CBCTs reconstructed by the conventional FDK algorithm were compromised by strong under-sampling artifacts, while the iterative ADMM algorithm with total variation regularization (ADMM-TV) manages to substantially suppress these artifacts but unable to reconstruct the anatomical details, especially for relatively small scan angles (30º). With diffusion models introduced as denoisers, DiffusionMBIR and DOLCE outperformed ADMM-TV, while the reconstructed anatomical details show clear mismatches with the 'ground-truth', especially for the 30º scan angle. In contrast, with the high-frequency prior CT condition and condition phasing-out schemes, PFGDM-A and PFGDM-B reconstructed the anatomical structures to match well with the 'ground-truth' for axial, coronal, and sagittal views, even for the 30º scan angle scenario. Their reconstruction results using single-view projections are generally better than DiffusionMBIR and DOLCE, even if DiffusionMBIR and DOLCE used the more efficient orthogonal-view sampling. PFGDM-A and PFGDM-B showed comparable performance under both single-view and orthogonal-view sampling conditions.

### 3.3 Quantitative evaluation

Table I quantitatively compares different methods on their reconstruction accuracy. Most methods used the orthogonal-view scheme (ortho), while PFGDM-A and PFGDM-B used both single-view (single) and orthogonal-view (ortho) sampling scenarios. As shown, PFGDM-A and PFGDM-B consistently outperformed other methods under all limited-angle scenarios. The advantage of PFGDM becomes especially evident when the scan angle decreases, showing the efficacy of introducing prior CT information when the on-board projection information is limited. For 90- and 120-degree reconstruction, PFGDM-B performed slightly better than PFGDM-A. Notably, PFGDM-A and PFGDM-B based on single-view projections already yielded better or comparable results when compared with other methods that were based on orthogonal-view projections.

**TABLE I.** Average (±s.d.) PSNR and SSIM results on the test set by different methods on axial slices. The arrows are pointing to the direction of improved accuracy. Best values and second-best values for each metric are color-coded.

| Metric | PSNR↑ | | | SSIM↑ | | |
|---|---|---|---|---|---|---|
| /Angle | 30° | 90° | 120° | 30° | 90° | 120° |
| FDK(Ortho) | 12.66±1.12 | 13.96±1.06 | 14.83±1.74 | 0.304±0.115 | 0.431±0.201 | 0.511±0.201 |
| ADMM-TV(Ortho) | 12.86±1.74 | 14.18±2.99 | 15.59±3.87 | 0.533±0.072 | 0.688±0.044 | 0.773±0.068 |
| DOLCE(Ortho) | 18.57±2.53 | 22.74±3.82 | 25.51±3.03 | 0.776±0.068 | 0.868±0.096 | 0.896±0.106 |
| DiffusionMBIR(Ortho) | 19.61±2.47 | 24.58±4.18 | 26.18±4.89 | 0.807±0.048 | 0.912±0.068 | 0.926±0.067 |
| PFGDM-B(Single) | 22.05±0.85 | 25.48±1.01 | 27.22±0.88 | 0.851±0.054 | 0.926±0.021 | 0.946±0.013 |
| PFGDM-B(Ortho) | 23.72±1.19 | **26.68±1.04** | **28.20±1.28** | 0.894±0.034 | **0.941±0.014** | **0.954±0.011** |
| PFGDM-A(Single) | 22.21±1.16 | 25.80±1.28 | 27.77±1.42 | 0.855±0.055 | 0.929±0.021 | 0.949±0.013 |
| PFGDM-A(Ortho) | **23.81±2.25** | 26.63±2.79 | 27.97±3.10 | **0.896±0.036** | 0.937±0.029 | 0.949±0.027 |

### 3.4 Extreme Cases

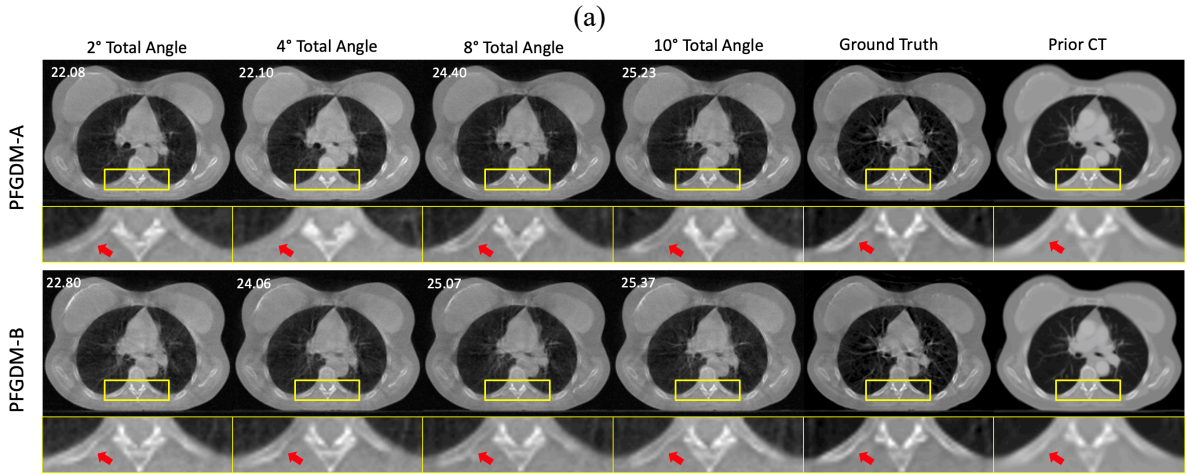

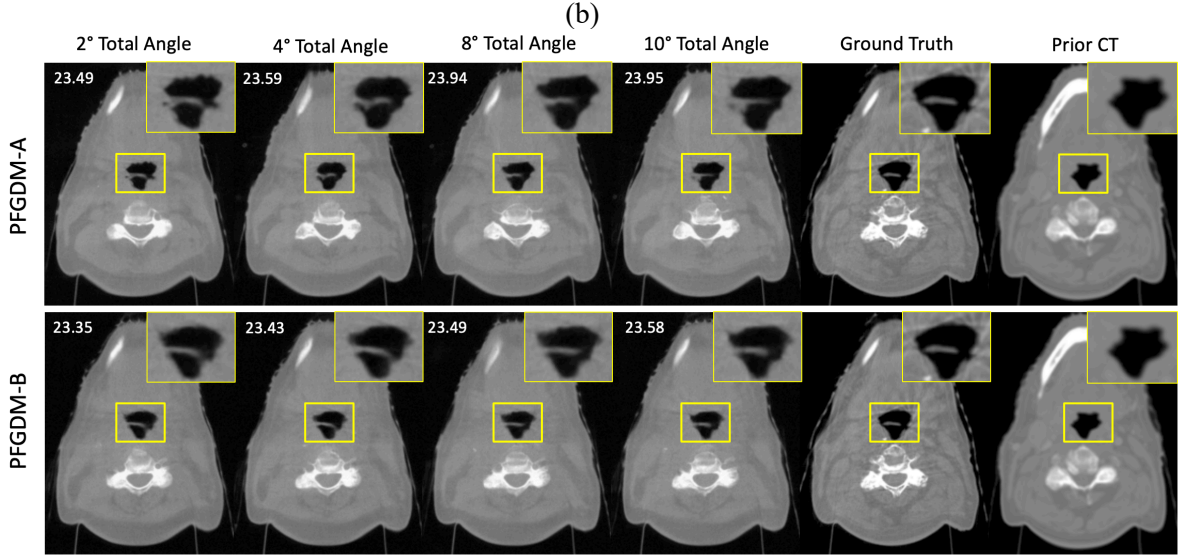

**Fig. 6.** Performance of PFGDM-A and PFGDM-B under extremely limited scan angle scenarios. The limited-angle projections were simulated with orthogonal-view acquisitions for a total scan angle of 2°([0°, 1°], [90°, 91°]), 4°([0°, 2°], [90°, 92°]), 8°([0°, 4°], [90°, 94°]), or 10°([0°, 5°], [90°, 95°]). PSNR scores of each reconstruction against the 'ground-truth' are indicated at the top-left corner of each figure. See Table II for quantitative results.

The visual comparison of the reconstructed LA-CBCTs of PFGDM-A and PFGDM-B using extremely limited scan angles is shown in Fig. 6. For Fig. 6 (a), PFGDM-B performed consistently better than PFGDM-A on this test sample with higher PSNR scores. PFGDM-B preserved the anatomical details of bony structures even though very limited information from CBCT projections is available. The slightly inferior performance of PFGDM-A could be due to the fact that the condition dropping step was optimized based on a different set of scan angles (Fig. 4). In Fig. 6 (b), PFGDM-A and PFGDM-B showed very similar performance. Even though the anatomical structures of the zoomed-in region showed clear differences between prior CT and 'ground-truth' CBCT, both of the PFGDM variants produced high-fidelity reconstructions of LA-CBCT. Overall, as shown in Table II, PFGDM-A and PFGDM-B had similar SSIM and PSNR scores.

**TABLE II.** Average (±s.d.) PSNR and SSIM results of PFGDM-A and PFGDM-B under extremely small limited-angle scenarios on axial slices. The arrows are pointing to the direction of improved accuracy.

| Metric /Angle | PSNR↑ | | | | SSIM↑ | | | |
|---|---|---|---|---|---|---|---|---|
| | 2° | 4° | 8° | 10° | 2° | 4° | 8° | 10° |
| PFGDM-B (Ortho) | **22.03±1.25** | **22.23±1.33** | 22.52±1.34 | 22.63±1.31 | 0.849±0.051 | 0.854±0.051 | 0.864±0.047 | 0.867±0.045 |
| PFGDM-A (Ortho) | 21.97±1.68 | 22.19±1.81 | **22.53±1.91** | **22.71±1.98** | **0.851±0.049** | **0.857±0.046** | **0.865±0.044** | **0.869±0.044** |

## 4. Discussion

CBCT plays a crucial role in image-guided radiotherapy, and there is a continuing interest in reconstructing CBCTs from limited-angle acquisitions (LA-CBCT) to enhance imaging efficiency, reduce radiation exposure, and improve mechanical clearance. LA-CBCT reconstruction, however, suffers from severe under-sampling artifacts and distortions, making it a highly ill-posed inverse problem. In this study, we proposed a diffusion model-based framework, prior frequency-guided diffusion model, to address the LA-CBCT reconstruction problem. PFGDM uses high-frequency information from patients' prior CT scans as anatomical detail-preserving conditions to feed into a diffusion model-based regularizer, which was incorporated into an iterative reconstruction framework based on the alternating direction method of multipliers (ADMM). The two implementations of this framework, PFGDM-A and PFGDM-B, yielded similar reconstruction results that were close to fully-sampled reconstructions, and outperformed standard FDK, regularized iterative reconstruction (ADMM-TV), and current SOTA diffusion network-based reconstructing approaches including DiffusionMBIR and DOLCE (Fig. 5 and Table I).

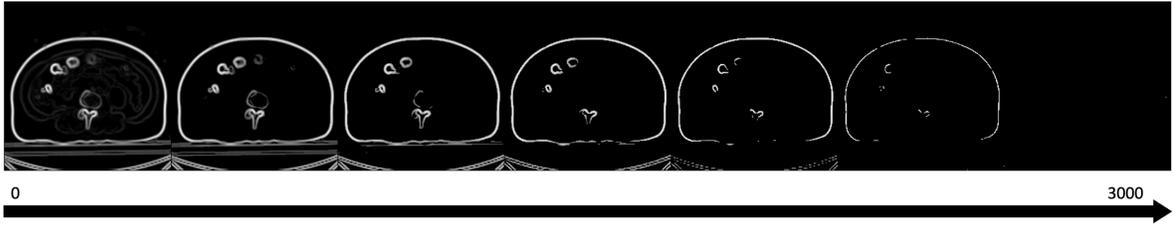

0                                                                                              3000

**Fig. 7.** Sample conditional input of PFGDM-B from reconstruction step 0 to step 3000, with step-size of 500.

Of the two PFGDM variants, PFGDM-A incorporates the same conditional input from the high-frequency information of the patient's prior CT until a pre-defined step. Our parametric study (Fig. 4) shows that discarding the conditional information at the intermediate step leads to better results, which stems from the anatomical changes between the planning CT and the daily CBCT (Fig. 3). Per our study, stopping introducing high-frequency conditions after 1400 reconstruction steps achieves the optimal result in general (Fig. 4). However, such an optimized number may change across different datasets or scanning conditions. Given a new data distribution, a parametric study may be needed to find the optimal condition dropping step. PFGDM-B, on the other hand, circumvents this potential issue by dynamically changing the conditional input during the entire reconstruction process. The high-frequency information corresponding to CT anatomical boundaries is gradually phased out to prevent the model from being overly influenced by the prior CT information (Fig. 7). In our experiments, the qualitative and quantitative results of PFGDM-A and PFGDM-B are comparable. However, in a few scenarios where CT and CBCT present clear anatomical differences, PFGDM-A was shown to provide better performance than PFGDM-B as the latter appeared more impacted by the incorrect condition (Fig. 8). More investigations are warranted in future studies to investigate the correlation between the condition's accuracy and the LA-CBCT reconstruction accuracy, as well as its impacts on the two PFGDM variants.

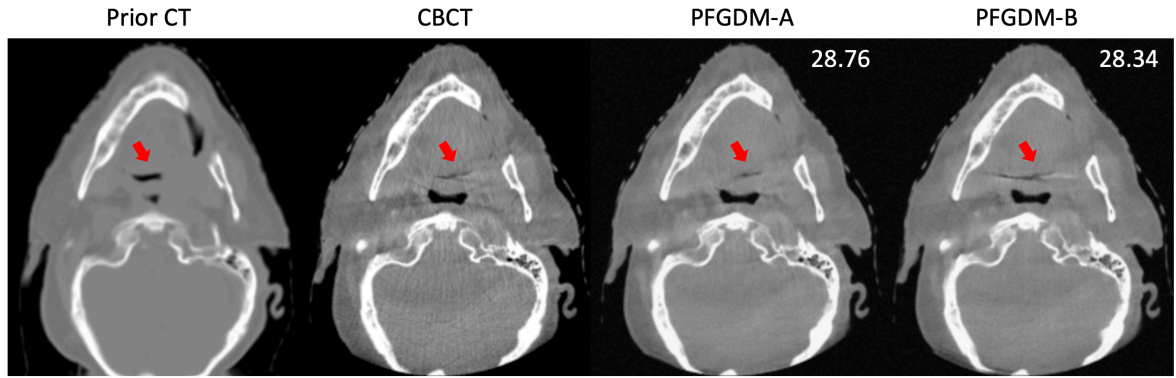

**Fig. 8.** PFGDM-A and PFGDM-B reconstruction results when CT/CBCT have clear anatomical differences. The limited-angle projections were simulated by the orthogonal-view geometry for a total scan angle of $120°([0°,60°],[90°,150°])$. PSNR scores of each reconstruction against the 'ground-truth' CBCT are indicated at the top-right corner.

The framework of PFGDM is built upon the assumption that patient-specific prior CTs share a similar underlying patient anatomy with the CBCTs. The assumption is generally valid although there exist various degrees of anatomical mismatches due to patient setup variations, anatomical deformations, treatment responses, or patient on-board motion. In this study, we pre-processed the CT and CBCT images by rigidly registering them, such that the impact of rigid patient setup variations can be reduced. In real clinical practices, a rigid setup correction can be performed to server the purpose via 2D radiography-based matching, or surface image-guided matching. However, the remaining deformation between CT and CBCT cannot be corrected and relies on the PFGDM to further reconstruct the anatomical differences. To alleviate the burden on PFGDM to reconstruct all the deformation-induced anatomical variations, we can potentially apply 2D–3D deformable registration (Zhang, 2021) to bring the prior CT closer to the to-be-solved CBCT, which may potentially further improve the accuracy of PFGDM. Alternatively, we can also use a fully-sampled CBCT volume acquired in previous treatment fractions to serve as the prior, which may contain a patient anatomy closer to the to-be-reconstructed CBCT and improve the overall reconstruction accuracy.

Our study further supported the finding that the orthogonal-view geometry yields better reconstruction results than the single-view geometry, even if the total scan angle remains the same. Therefore, orthogonal-view projections are preferred in limited-angle CBCT reconstruction tasks. In our study the orthogonal-view projections were simulated in kilovoltage (kV) energy for both directions. However, simultaneous orthogonal-view kV–kV acquisitions are currently not available on clinical radiotherapy devices and limited to experimental setups only (Ren *et al.*, 2016). With orthogonally-arranged kV and megavoltage (MV) imaging systems equipped on most modern LINACs, orthogonal-view kV–MV projections are available for PFGDM reconstruction. Previous studies have attempted to harmonize the intensity differences between the kV and MV projections by applying pixel intensity linear scaling or using normalized cross correlation as the similarity metric (Ren *et al.*, 2014; Zhang *et al.*, 2017b), which however requires further investigations for potential translations into PFGDM reconstruction.

The study shows the feasibility of reconstructing CBCT images from limited-angle 2D X-ray projections guided by a conditional diffusion model. However, several limitations need to be considered and addressed before potential clinical applications. First, the reconstruction of PFGDM is time consuming, which takes around 3 hours to reconstruct a CBCT volume of 80 slices. The low efficiency is mainly caused by the slow diffusion models. To improve the efficiency, the current score-based diffusion model can be replaced by the DDIM (Denoising Diffusion Implicit Model) (Song *et al.*, 2020a) which allows step-skipping in the denoising process for acceleration. In addition, further parallelization of the ADMM optimization can also boost the reconstruction speed (Deng *et al.*, 2017). Second, the cone-beam X-ray projections were simulated in our experiments from CBCT images. Further evaluation of PFGDM on real cone-beam projections is warranted. Third, we used fully-sampled FDK CBCTs from the clinic as the 'ground-truth' in this study. However, these images still contain artifacts (Schulze *et al.*, 2011) that compromise the fidelity of the 'ground-truth', leading to potential inaccuracies in quantitative analysis. As an alternative, techniques such as iterative reconstruction (Sun *et al.*, 2015) and deep learning-based methods (Bayaraa *et al.*, 2020; Amirian *et al.*, 2024) have shown potential in enhancing CBCT quality and reducing artifacts, which can potentially be used to reconstruct the CBCTs to serve as the 'ground-truth', if the raw X-ray projections are available.

## 5. Conclusion

In this work, we introduced PFGDM, a diffusion model-based method for LA-CBCT reconstruction. PFGDM incorporates the high-frequency information from the patient's existing CT scans as additional priors to compensate the missing angle information in LA-CBCT reconstruction. Within this framework, we further developed two variants, PFGDM-A and PFGDM-B, to evaluate different condition phasing-out strategies to allow the PFGDM to better reconstruct CT/CBCT differences. Both variants of PFGDM outperformed the other LA-CBCT reconstruction techniques, especially for relatively small scan angles. The tests on extremely limited scan angles also demonstrated the potential of PFGDM-based methods. PFGDM makes a general method that does not need re-training for different imaging geometries or limited-angle scenarios, providing it a substantial advantage over methods like DOLCE for potential clinical translation.

## Acknowledgments

The study was supported by the US National Institutes of Health (R01 CA240808, R01 CA258987, and R01 CA280135).

## Conflict of interest statement

The authors have no relevant conflicts of interest to disclose.

## Ethical statement

The datasets used in this study were retrospectively collected from an approved study at UT Southwestern Medical Center on August 31, 2023, under an umbrella IRB protocol 082013-008 (Improving radiation treatment quality and safety by retrospective data analysis). This is a retrospective analysis study and not a clinical trial. No clinical trial ID number is available. Individual patient consent was signed for the anonymized use of the imaging and treatment planning data for retrospective analysis. These studies were conducted in accordance with the principles embodied in the Declaration of Helsinki.

# References

Amirian M, Barco D, Herzig I and Schilling F P 2024 Artifact Reduction in 3D and 4D Cone-Beam Computed Tomography Images With Deep Learning: A Review *Ieee Access* **12** 10281-95

Anderson B D O 1982 Reverse-time diffusion equation models *Stochastic Processes and their Applications* **12** 313-26

Barutcu S, Aslan S, Katsaggelos A K and Gürsoy D 2021 Limited-angle computed tomography with deep image and physics priors *Sci Rep-Uk* **11**

Bayaraa T, Hyun C M, Jang T J, Lee S M and Seo J K 2020 A Two-Stage Approach for Beam Hardening Artifact Reduction in Low-Dose Dental CBCT *Ieee Access* **8** 225981-94

Chang W, Lee J M, Lee K, Yoon J H, Yu M H, Han J K and Choi B I 2013 Assessment of a Model-Based, Iterative Reconstruction Algorithm (MBIR) Regarding Image Quality and Dose Reduction in Liver Computed Tomography *Invest Radiol* **48** 598-606

Chen C Y, Xing Y X, Gao H W, Zhang L and Chen Z Q 2022 Sam's Net: A Self-Augmented Multistage Deep-Learning Network for End-to-End Reconstruction of Limited Angle CT *Ieee T Med Imaging* **41** 2912-24

Chen H, Zhang Y, Kalra M K, Lin F, Chen Y, Liao P X, Zhou J L and Wang G 2017 Low-Dose CT With a Residual Encoder-Decoder Convolutional Neural Network *Ieee T Med Imaging* **36** 2524-35

Chung H, Ryu D, Mccann M T, Klasky M L and Ye J C 2023 Solving 3D Inverse Problems using Pre-trained 2D Diffusion Models *2023 Ieee/Cvf Conference on Computer Vision and Pattern Recognition (Cvpr)* 22542-51

Chung H, Sim B and Ye J C 2022 Come-Closer-Diffuse-Faster: Accelerating Conditional Diffusion Models for Inverse Problems through Stochastic Contraction *2022 Ieee/Cvf Conference on Computer Vision and Pattern Recognition (Cvpr)* 12403-12

Chung H Y J and Ye J C 2022 Score-based diffusion models for accelerated MRI *Med Image Anal* **80**

Deng W, Lai M J, Peng Z M and Yin W T 2017 Parallel Multi-Block ADMM with

$(1 /$

$)$ Convergence *J Sci Comput* **71** 712-36

Dhariwal P and Nichol A 2021 Diffusion Models Beat GANs on Image Synthesis *Adv Neur In* **34**

Feldkamp L A, Davis L C and Kress J W 1984 Practical Cone-Beam Algorithm *J Opt Soc Am A* **1** 612-9

Gao Q F, Ding R, Wang L Y, Xue B and Duan Y P 2023 LRIP-Net: Low-Resolution Image Prior-Based Network for Limited-Angle CT Reconstruction *Ieee T Radiat Plasma* **7** 163-74

Gu J and Ye J C 2017 Multi-Scale Wavelet Domain Residual Learning for Limited-Angle CT Reconstruction *ArXiv* **abs/1703.01382**

Gupta H, Jin K H, Nguyen H Q, McCann M T and Unser M 2018 CNN-Based Projected Gradient Descent for Consistent CT Image Reconstruction *Ieee T Med Imaging* **37** 1440-53

Ho J, Jain A and Abbeel P 2020 Denoising Diffusion Probabilistic Models *ArXiv* **abs/2006.11239**

Hu D L, Zhang Y K, Liu J, Luo S H and Chen Y 2022 DIOR: Deep Iterative Optimization-Based Residual-Learning for Limited-Angle CT Reconstruction *Ieee T Med Imaging* **41** 1778-90

Je U, Cho H, Lee M, Oh J, Park Y, Hong D, Park C, Cho H, Choi S and Koo Y 2014 Dental cone-beam CT reconstruction from limited-angle view data based on compressed-sensing (CS) theory for fast, low-dose X-ray imaging *J Korean Phys Soc* **64** 1907-11

Kanopoulos N, Vasanthavada N and Baker R L 1988 Design of an Image Edge-Detection Filter Using the Sobel Operator *Ieee J Solid-St Circ* **23** 358-67

Katsura M, Matsuda I, Akahane M, Sato J, Akai H, Yasaka K, Kunimatsu A and Ohtomo K 2012 Model-based iterative reconstruction technique for radiation dose reduction in chest CT: comparison with the adaptive statistical iterative reconstruction technique *Eur Radiol* **22** 1613-23

Klein S, Staring M, Murphy K, Viergever M A and Pluim J P W 2010 elastix: A Toolbox for Intensity-Based Medical Image Registration *Ieee T Med Imaging* **29** 196-205

Kohr H and Adler J 2017 *ODL (Operator Discretization Library)*

Li Y, Shao H-C, Liang X, Chen L, Li R, Jiang S B, Wang J and Zhang Y 2023 Zero-Shot Medical Image Translation via Frequency-Guided Diffusion Models *Ieee T Med Imaging* **43** 980-93

Li Z H, Zhang W K, Wang L Y, Cai A L, Liang N N, Yan B and Li L 2019 A Sinogram Inpainting Method based on Generative Adversarial Network for Limited-angle Computed Tomography *Proc Spie* **11072**

Lin W A, Liao H F, Peng C, Sun X H, Zhang J D, Luo J B, Chellappa R and Zhou S K 2019 DuDoNet: Dual Domain Network for CT Metal Artifact Reduction *Proc Cvpr Ieee* 10504-13

Liu J, Anirudh R, J. Thiagarajan J, He S, Mohan K A, Kamilov U and Kim H 2022 *DOLCE: A Model-Based Probabilistic Diffusion Framework for Limited-Angle CT Reconstruction*

Liu L 2014 Model-based Iterative Reconstruction: A Promising Algorithm for Today's Computed Tomography Imaging *J Med Imaging Radiat* **45** 131-6

Pickhardt P J, Lubner M G, Kim D H, Tang J, Ruma J A, del Rio A M and Chen G H 2012 Abdominal CT With Model-Based Iterative Reconstruction (MBIR): Initial Results of a Prospective Trial Comparing Ultralow-Dose With Standard-Dose Imaging *Am J Roentgenol* **199** 1266-74

Ren L, Chen Y X, Zhang Y, Giles W, Jin J Y and Yin F F 2016 Scatter Reduction and Correction for Dual-Source Cone-Beam CT Using Prepatient Grids *Technol Cancer Res T* **15** 416-27

Ren L, Zhang Y and Yin F F 2014 A limited-angle intrafraction verification (LIVE) system for radiation therapy *Med Phys* **41**

Rui X, Cheng L S, Long Y, Fu L, Alessio A M, Asma E, Kinahan P E and De Man B 2015 Ultra-low dose CT attenuation correction for PET/CT: analysis of sparse view data acquisition and reconstruction algorithms *Phys Med Biol* **60** 7437-60

Särkkä S and Solin A 2019 *Applied Stochastic Differential Equations* vol 10 (Cambridge: Cambridge University Press)

Schulze R, Heil U, Gross D, Bruellmann D D, Dranischnikow E, Schwanecke U and Schoemer E 2011 Artefacts in CBCT: a review *Dentomaxillofac Rad* **40** 265-73

Sidky E Y and Pan X C 2008 Image reconstruction in circular cone-beam computed tomography by constrained, total-variation minimization *Phys Med Biol* **53** 4777-807

Song J, Meng C and Ermon S 2020a Denoising Diffusion Implicit Models *ArXiv* **abs/2010.02502**

Song Y and Ermon S 2019 Generative Modeling by Estimating Gradients of the Data Distribution *Advances in Neural Information Processing Systems 32 (Nips 2019)* **32**

Song Y and Ermon S 2020 Improved Techniques for Training Score-Based Generative Models *ArXiv* **abs/2006.09011**

Song Y, Shen L, Xing L and Ermon S 2021 Solving Inverse Problems in Medical Imaging with Score-Based Generative Models *ArXiv* **abs/2111.08005**

Song Y, Sohl-Dickstein J N, Kingma D P, Kumar A, Ermon S and Poole B 2020b Score-Based Generative Modeling through Stochastic Differential Equations *ArXiv* **abs/2011.13456**

Sun T, Sun N B, Wang J and Tan S 2015 Iterative CBCT reconstruction using Hessian penalty *Phys Med Biol* **60** 1965-87

Vincent P 2011 A Connection Between Score Matching and Denoising Autoencoders *Neural Comput* **23** 1661-74

Virtanen P, Gommers R, Oliphant T E, Haberland M, Reddy T, Cournapeau D, Burovski E, Peterson P, Weckesser W, Bright J, van der Walt S J, Brett M, Wilson J, Millman K J, Mayorov N, Nelson A R J, Jones E, Kern R, Larson E, Carey C J, Polat I, Feng Y, Moore E W, VanderPlas J, Laxalde D, Perktold

J, Cimrman R, Henriksen I, Quintero E A, Harris C R, Archibald A M, Ribeiro A N H, Pedregosa F, van Mulbregt P and Contributors S 2020 SciPy 1.0: fundamental algorithms for scientific computing in Python (vol 33, pg 219, 2020) *Nat Methods* **17** 352-

Walter C, Schmidt J C, Rinne C A, Mendes S, Dula K and Sculean A 2020 Cone beam computed tomography (CBCT) for diagnosis and treatment planning in periodontology: systematic review update *Clin Oral Invest* **24** 2943-58

Wang Z, Bovik A C, Sheikh H R and Simoncelli E P 2004 Image quality assessment: From error visibility to structural similarity *Ieee T Image Process* **13** 600-12

Yang X, Kahnt M, Brückner D, Schropp A, Fam Y, Becher J, Grunwaldt J D, Sheppard T L and Schroer C G 2020 Tomographic reconstruction with a generative adversarial network *J Synchrotron Radiat* **27** 486-93

Zhang Y 2021 An unsupervised 2D-3D deformable registration network (2D3D-RegNet) for cone-beam CT estimation *Phys Med Biol* **66**

Zhang Y, Ma J H, Iyengar P, Zhong Y C and Wang J 2017a A new CT reconstruction technique using adaptive deformation recovery and intensity correction (ADRIC) *Med Phys* **44** 2223-41

Zhang Y, Ren L, Ling C C and Yin F F 2013a Respiration-phase-matched digital tomosynthesis imaging for moving target verification: a feasibility study *Med Phys* **40** 071723

Zhang Y, Ren L, Vergalasova I and Yin F F 2017b Clinical Study of Orthogonal-View Phase-Matched Digital Tomosynthesis for Lung Tumor Localization *Technol Cancer Res Treat* **16** 866-78

Zhang Y, Yin F-F, Pan T, Vergalasova I and Ren L 2015a Preliminary clinical evaluation of a 4D-CBCT estimation technique using prior information and limited-angle projections *Radiotherapy and Oncology* **115** 22-9

Zhang Y, Yin F F and Ren L 2015b Dosimetric verification of lung cancer treatment using the CBCTs estimated from limited-angle on-board projections *Med Phys* **42** 4783-95

Zhang Y, Yin F F, Segars W P and Ren L 2013b A technique for estimating 4D-CBCT using prior knowledge and limited-angle projections *Med Phys* **40**

Zhou B, Zhou S K, Duncan J S and Liu C 2021 Limited View Tomographic Reconstruction Using a Cascaded Residual Dense Spatial-Channel Attention Network With Projection Data Fidelity Layer *Ieee T Med Imaging* **40** 1792-804